\documentclass[prl,twocolumn,showpacs,preprintnumbers,amsmath,amssymb]{revtex4}

\usepackage{dcolumn}% Align table columns on decimal point
\usepackage{bm}% bold math

\begin{document}

\title{Information-theoretic natural ultraviolet cutoff for spacetime}
%\title{Spacetime can be simultaneously discrete and continuous, in the
%same way that information can}

\author{Achim Kempf}
\affiliation{
Departments of Applied Mathematics and Physics, University of Waterloo\\
Waterloo, Ontario N2L 3G1, Canada}

%\date{\today}

\begin{abstract}
Fields in spacetime could be simultaneously discrete \it and \rm continuous, in
the same way that information can: it has been shown that the amplitudes,
$\phi(x_n)$, that a field takes at a generic discrete set of points, $x_n$, can
be sufficient to reconstruct the field $\phi(x)$ for all $x$, namely if there
exists a certain type of natural ultraviolet (UV) cutoff in nature, and if the
average spacing of the sample points is at the UV cutoff scale. Here, we
generalize this information-theoretic framework to spacetimes themselves. We
show that samples taken at a generic discrete set of points of a
Euclidean-signature spacetime can allow one to reconstruct the shape of that
spacetime everywhere, down to the cutoff scale. The resulting methods could be
useful in various approaches to quantum gravity.

\end{abstract}


\maketitle

At the heart of any candidate theory of quantum gravity is an attempt to
describe the structure of spacetime at the Planck scale. The problem is hard
because general relativity and quantum theory provide seemingly contradictory
indications. While general relativity describes spacetime as a manifold,
quantum field theories appear to be well-defined only if spacetime is discrete.

In this context, it has been proposed that spacetime could be simultaneously
discrete and continuous, in the same way that information can
\cite{ak-prl-2000}. The underlying mathematical structure, Shannon sampling
theory \cite{shannon}, is at the heart of information theory and is of
ubiquitous use in communication engineering and signal processing. Sampling
theory explains that any signal, $\phi(t)$, with a bandlimit, $\Omega$, can be
reconstructed perfectly for all $t$ from the knowledge of the samples
$\phi(t_n)$ at sample points $t_n$ whose average density (technically, the
Beurling density) is at least $2\Omega$. The reconstruction of $\phi(t)$ is
numerically most stable when the samples are taken equidistantly, with spacing
$t_{n+1}-t_n=(2\Omega)^{-1}$, in which case the reconstruction formula reads:
$\phi(t) = \sum_n \mbox{sinc}(2(t-t_n)\Omega)~\phi(t_n)$. For later reference,
we note that if $\phi$ and $\psi$ are bandlimited then also,
$\sum_n\phi(t_n)\psi(t_n)=2\Omega\int dt~\phi(t)\psi(t)$, which is useful, \it
e.g., \rm to solve hard-to-sum series ({\it e.g.}, in analytic number theory):
view the terms of the series in question as samples of bandlimited functions,
rewrite the series as an integral, then apply integration tools such as
integration by parts or contour integration.

The idea that physical fields could possess the sampling property was first
proposed in \cite{ak-prl-2000}, where it was shown that fields possess the
sampling property if the uncertainty relations are modified in the ultraviolet
so that there is formally a finite lower bound, $\Delta x_{min}$, on the
uncertainty in position. Such uncertainty relations have indeed arisen in
various studies of quantum gravity and string theory, see {\it e.g.},
\cite{ucrs}. This UV cutoff has then been applied, in particular, to the
space-like hypersurfaces in inflationary cosmology. Possible signatures in the
scalar and tensor spectra of the cosmic microwave background have been
calculated, see {\it e.g.} \cite{cosmo}.

\bf Sampling theory of fields. \rm Let us briefly review the generalization of
sampling theory to physical fields in curved Euclidean-signature spacetimes,
such as the fields that are being summed over in the path integral of Euclidean
quantum field theory (QFT), \cite{ak-prl-2004,ak-prl-2008}.

To this end, consider a spacetime described by a compact smooth Riemannian
manifold. For simplicity, let us assume it has no boundary. For the covariant
inner product of fields on the manifold we use the usual bra-ket notation
inspired by first quantization: $(\phi\vert\psi)=\int d^dx~\sqrt{\vert
g\vert}~\phi(x)\psi(x)$, so that one has, for example: $\phi(x)= (x\vert\phi)$.
We use the sign convention in which the spectrum of the Laplacian is positive
and we choose $c=\hbar=G=1$. The Laplacian is self-adjoint, with $\Delta
v_{\lambda_i}=\lambda_i v_{\lambda_i}$ solved by normalizable eigenfunctions
with discrete eigenvalues. The generalization of the assumption of
bandlimitation is the assumption that there exists a natural hard UV cutoff,
$\Lambda$, of the spectrum of the Laplacian, with $\Lambda$, for example, at
the Planck scale. In QFT, the space of fields, ${\cal F}$, that is being
integrated over in the path integral, is then spanned by those eigenfunctions
$v_{\lambda_i}$ of the Laplacian whose eigenvalues, $\lambda_i$, are below the
cutoff, $\lambda_i <\Lambda$. Let $P$ denote the projector onto ${\cal F}$ and
let us denote the Laplacian restricted to ${\cal F}$ by $\Delta_c = P\Delta P$.
The fields $\vert\phi)\in{\cal F}$ that occur in the path integral obey
$\phi(x)= (x\vert\phi)=(x\vert P\vert \phi)$. Notice that this means that the
point-localized fields $\vert x )$ are now indistinguishable from the fields
$P\vert x )$ in which wavelengths shorter than the cutoff scale are removed.
Intuitively, this expresses a minimum length uncertainty principle.

Since the Laplacian's spectrum does not possess accumulation points, the
dimension, $N$, of the space of fields is finite, dim$({\cal F})=N$. It was
shown that therefore any field $\phi(x)\in{\cal F}$ can be reconstructed
everywhere if known only on $N$ generic points of the manifold. The fields,
actions and equations of motion therefore possess a representation on the
smooth spacetime manifold as well as equivalently also on any lattice of $N$
generic points. Under mild conditions, Weyl's asymptotic formula was shown to
imply that as the infrared (IR) cutoff is removed by letting the volume of the
manifold diverge, $V\rightarrow\infty$, one has $N\rightarrow \infty$ such that
the density of samples necessary for reconstruction, $N/V$, indeed stays finite
\cite{ak-prl-2008}.
%\smallskip\newline

\bf Sampling theory of spacetime. \rm
%\smallskip\newline
The aim in this Letter is to generalize sampling theory to spacetime itself.

Assuming the natural hard UV cutoff above, can the shape (curvature and global
topology) of a Euclidean-signature spacetime be reconstructed everywhere from
suitable samples taken at a discrete set of points?

For most purposes, a spacetime's shape is best described in terms of the affine
connection. Here, however, a different description appears more useful. Let us
recall a comment by Einstein \cite{Einstein}, who pointed out that the
nontrivial shape of a manifold manifests itself not only in the nontriviality
of the parallel transport of tensors. Crediting Helmholtz, Einstein emphasized
that the shape of a manifold can also be thought of in terms of the
nontriviality of the mutual distances among points: In $d$-dimensional flat
space, consider $M$ points. In cartesian coordinates, the points possess $M d$
coordinates $x^{(n)}_i$ with $n=1,...,M$ and $i=1,...,d$. By Pythagoras, the
$M(M-1)/2$ mutual distances $s_{n,n'}$ obey the equations $s_{n,n'}^2 =
\sum_{i=1}^d (x^{(n)}_i-x^{(n')}_i)^2$. If $M>2d+1$, the $M d$ coordinates can
be eliminated in these $M(M-1)/2$ equations, to leave $M(M-1)/2 -M d$
nontrivial equations that must hold among the mutual distances $s_{n,n'}$ if
the manifold is indeed flat. If the manifold is curved this manifests itself in
the way in which these equations are violated.

Let us try, therefore, to reconstruct the shape of a spacetime of finite volume
by sampling at a sufficient number, $N$, of generic points a quantity that is
closely related to their mutual distances. To this end, we sample the
propagator, or correlator, $G(x^{(n)},x^{(n')})$, of a scalar field for each
pair of the $N$ chosen points. Generally, the larger the distance between
$x^{(n)}$ and $x^{(n')}$, the smaller is the correlator. For a free scalar we
have, {\it e.g.}: $G(x^{(n)},x^{(n')}) = (x^{(n)} \vert P (\Delta +m^2)^{-1}
P\vert x^{(n')} )$. Indeed, the knowledge of the $N(N-1)/2$ matrix elements
$G(x^{(n)},x^{(n')})$ suffices to reconstruct the shape of the spacetime up to
the UV cutoff scale. To see this, we note first that the matrix
$(G(x^{(n)},x^{(n')}))_{nn'}$ represents the correlator, $(\Delta_c
+m^2)^{-1}$, in a basis, namely the basis $\{P\vert x^{(n)} )\}$, which means
that we can determine its eigenvalues. Since the correlator is diagonal in the
same basis as the Laplacian $\Delta_c$, we also obtain the spectrum of
$\Delta_c$.
%(Also in interacting theories the Laplacian's spectrum can be
%calculated since the functional dependence of the correlator on $\Delta_c$ is
%still determined by the action.)

Crucially now, the eigenvalues of $\Delta_c$ provide us with the shape of the
spacetime from large length scales down to the cutoff scale. To see this, let
us recall key results of the discipline of spectral geometry which studies the
relationship between the shape of a manifold (or a domain) and the spectrum of
its Laplacian or Dirac operator. (Note that spectral geometry thereby naturally
combines functional analysis and differential geometry, {\it i.e.}, the
mathematical languages of quantum theory and general relativity.) In
particular, isospectral manifolds are generally also isometric, though there
are exceptions. One cannot always ``hear the shape of a drum", see, {\it e.g.},
\cite{Gordon}. Even pairs of isospectral but non-isometric manifolds that are
compact and simply connected have been constructed \cite{Schueth}. It is known,
however, that the eigenvalues change continuously as a function of the shape of
the manifold. Also, the eigenvalues are nondegenerate for generic manifolds,
and a manifold can have degenerate eigenvalues only if it possesses a
continuous group of isometries \cite{Bando}.

For our purposes, we are led to consider classes of manifolds whose Laplacians
share the same eigenvalues (and their multiplicities) only up to the cutoff
$\Lambda$, and which we may therefore call $\Lambda$-isospectral. This is
because the samples of the matrix elements of the correlator,
$(G(x^{(n)},x^{(n')}))_{nn'}$, determine only the $N$ eigenvalues of the
Laplacian $\Delta_c$. The eigenvalues that the full Laplacian, $\Delta$,
possesses beyond the cutoff remain undetermined. This tells us what the UV
cutoff means for the shape of spacetime itself. The cutoff does not directly
mean a cutoff for the curvature, for example. Instead, the fact that the
eigenvalues of the Laplacian beyond the cutoff remain undetermined by any
measurement possible means that all $\Lambda$-isospectral manifolds are
physically indistinguishable and thus equivalent. When referring to a
``spacetime with UV cutoff", specified by $N$ eigenvalues, $\lambda_1 \le
...\le \lambda_N<\Lambda$, we will therefore henceforth mean an equivalence
class of $\Lambda$-isospectral manifolds.

Intuitively, the higher the eigenvalues of the Laplacian the higher the
``squared momentum" that they represent, and therefore the smaller the wrinkles
which they determine in the manifold. The eigenvalues up to $\Lambda$
essentially determine the shape of a spacetime with UV-cutoff from large scales
down to lengths as small as the cutoff scale. The undetermined eigenvalues
beyond the cutoff would describe wrinkles on length scales smaller than the
cutoff scale. At distances smaller than the cutoff scale the shape of a
spacetime with UV cutoff is not determined.

This can be viewed as a matter of representation theory. In general relativity,
the choice of coordinate system is merely a choice of representation for an
underlying Riemannian manifold. With the UV cutoff, even the choice of
Riemannian manifold is merely a choice of representation for an underlying
``spacetime with UV cutoff" that is defined through the first $N$ eigenvalues
of the Laplacian.

To be precise, however, the picture is slightly more subtle. A theory may
contain additional fields with interactions that allow one to physically
distinguish among certain $\Lambda$-isospectral manifolds. At the very least,
we have to divide each equivalence class of $\Lambda$-isospectral manifolds
into sub-equivalence classes of manifolds that are continuously deformable into
each other within their class of $\Lambda$-isospectral manifolds. This is
because at least for those subclasses, and possibly also for sub-subclasses
within them, we can define what we may call ``geometric quantum numbers" that
distinguish them and that could be measurable in the full theory. Consider, for
example, a manifold in the shape of a potato's surface, with N points singled
out. The same N points on the manifold with the shape of the mirror-imaged
potato clearly possess the same correlators. Thus, the two potato surfaces,
similar to enantiomeres in chemistry, are $\Lambda$-isospectral. However, if
the action contains a parity-breaking interaction, such as the weak
interaction, then the two manifolds become distinguishable. Apart from parity,
also the dimension of the manifold can be viewed as a geometric quantum number,
measurable, \it e.g., \rm through interactions involving tensors (whose
dimensions indicate the manifold's dimension). Indeed, $\Lambda$-isospectral
manifolds of different dimensions cannot be continuously deformed into another
\cite{Bando}. Note that the scaling of the Laplacian's spectrum for
asymptotically large eigenvalues is in one-to-one correspondence to the
manifold's dimension \cite{piiss}. It should be interesting to study the set of
possible geometric quantum numbers, and their relation to cohomology, by
methods similar to those used to construct isospectral non-isometric manifolds,
see {\it e.g.} \cite{Schueth}.

A ``spacetime with UV cutoff" is, therefore, an equivalence class of manifolds
that are $\Lambda$-isospectral and possess the same geometric quantum numbers.
An intriguing possibility is that it may not be necessary to keep track of
geometric quantum numbers as variables that are separate from the spectrum
after all, namely when working with the full Laplacian, $d\delta+\delta d$, on
all differential forms, and the Dirac operator. Their spectra may well include
all information about the geometric quantum numbers. It is known, for example,
that the multiplicity of the eigenvalue $0$ of the Laplace operator on
$p$-forms coincides with the $p$'th Betti number of the manifold \cite{piiss},
and that the largest value of $p$ is the manifold's dimension.

Let us summarize the sampling and reconstruction for both spacetime and field.
Abstractly, a spacetime is specified by
%geometric quantum numbers and
the eigenvalues $\lambda_1,...,\lambda_N$ of $\Delta_c$. A field, $\vert \phi
)$, on the spacetime is a vector in the $N$-dimensional Hilbert space on which
$\Delta_c$ acts, conveniently specified through its coefficients $\phi_i$ in an
ON eigenbasis $\{\vert v_{\lambda_i} )\}$ of $\Delta_c$, as $\vert\phi )=
\sum_{i=1}^N \phi_i \vert v_{\lambda_i} )$. Now consider a continuous
representation of the spacetime as a manifold that possesses the right
geometric quantum numbers, such as the dimension, and whose Laplacian $\Delta$
possesses the spectrum of $\Delta_c$ up to $\Lambda$. Choosing coordinates, the
eigenvectors of $\Delta_c$ are represented as eigenfunctions $v_{\lambda_i}(x)$
of $\Delta$. The field is represented by the function
$\phi(x)=\sum_{i=1}^N\phi_i v_{\lambda_i}(x)$. We obtain a lattice
representation by choosing $N$ generic points $x^{(1)},...,x^{(N)}$ and then
sampling the matrix of correlators $G(x^{(i)},x^{(j)})$ and the field's
amplitudes $\phi(x^{(i)})$. (Remark: if the theory has only one field we also
record the overlap matrix $B_{ij}=(x^{(i)}\vert P\vert x^{(j)})$. This is not
necessary in general, \it i.e., \rm $B$ is is implied, if the theory has
multiple fields, such as, \it e.g., \rm two scalar fields with different
masses.) From these data we can fully reconstruct the $\lambda_1,...,\lambda_N$
and $\phi_1,...,\phi_N$ that abstractly define the spacetime and field. First,
the diagonalization of the matrix of correlators yields the eigenvalues of
$\Delta_c$. To obtain the coefficients $\phi_j=( v_{\lambda_j}\vert\phi )$ we
insert a resolution of the identity in $( x^{(n)} \vert \phi  ) = \sum_{j=1}^N
( x^{(n)}\vert v_{\lambda_j} ) ~( v_{\lambda_j} \vert \phi  )$, {\it i.e.},
$\phi(x^{(n)})=\sum_j E_{nj}~\phi_j$. The change of basis matrix $E_{nj}=(
x^{(n)}\vert v_{\lambda_j} )$ is known from the diagonalization of the matrix
of correlator samples. Since the $\phi(x^{(i)})$ are known samples, we obtain
$\phi_i=\sum_j (E_{ij})^{-1}~\phi(x^{(j)})$. Thus, from the samples of field
amplitudes and correlators, we have obtained the spacetime in terms of the
eigenvalues of its Laplacian $\Delta_c$ and the field as a vector in the
$N$-dimensional vector space on which $\Delta_c$ acts. At this point we are
free to represent the abstract spacetime by the same or any other member of its
equivalence class of $\Lambda$-isospectral manifolds, choose coordinates and
express the field as an explicit function. Note that the choice of real-valued
orthonormalized eigenfunctions, while normally ambiguous up to a factor of
$-1$, is here fixed by continuity. The formalism establishes, therefore, an
equivalence between discrete and continuous representations of spacetimes and
fields.

Note that the reconstruction of the spacetime and the field is possible even
when the sample points $x_n$ are chosen close together, possibly leaving large
regions without samples. As mentioned, also in Shannon sampling the
reconstruction from irregularly-spaced samples is possible but requires
increased numerical precision. This can be understood by considering Shannon's
channel capacity formula, $C=\Omega \log_2(1+S/N)$ and the phenomenon of
superoscillations \cite{shannon}. While holding the overall information
density, $C$, fixed, the sample density, or Nyquist rate, $2\Omega$, can be
locally lowered, at the cost of needing an exponentially higher signal to noise
ratio $S/N$, which represents the reconstruction instability. In our case here,
we notice that if the $x_n$ are chosen close, the matrix $B$ shows that the
basis vectors $P\vert x_n )$ acquire significant overlap and thus form small
angles in the Hilbert space ${\cal F}$. This means that the condition number of
the change of basis matrix $E$ to the ON eigenbasis of $\Delta_c$,
deteriorates, which indicates the need for increased numerical precision. It
should be very interesting to develop the corresponding classical and quantum
capacity formulas for the sampling of manifolds.
\smallskip\newline
\bf A simple example of a partition function. \rm
%\smallskip\newline
Consider a matter action of the form $S_{matter} = \int d^nx~\sqrt{\vert
g\vert}~\left(\frac{1}{2}\phi(x)(\Delta+m^2)\phi(x)+J(x)\Phi(x)\right)$ where
we made a scalar field explicit and where $J$ can be a source or stand for a
background of other fields. Representation independently, the action reads:
$S_{matter} =\frac{1}{2} (  \phi\vert \Delta+m^2\vert\phi ) +(  J\vert \phi ) =
Tr\left(\frac{1}{2}(\Delta+m^2) \vert\phi )( \phi\vert +\vert J )(
\phi\vert\right)$. Representing the action in an ON eigenbasis of the
Laplacian, we have: $S_{matter} = \sum_{i=1}^N
\frac{1}{2}\phi_i(\lambda_i+m^2)\phi_i+ J_i\phi_i$.

The action already contains the degrees of freedom, $\lambda_i$, which describe
the shape of the spacetime. However, an action is still needed for the degree
of freedom, $N$, which describes the overall size of the system. The simplest
choice is $S_{size}=\alpha N = \alpha Tr(1)$, where $\alpha$ remains to be
determined. We obtain: $S_{total} =Tr\left(\alpha ~1+\frac{1}{2}(\Delta+m^2)
\vert\phi )( \phi\vert +\vert J )( \phi\vert\right)$. To understand $S_{size}$,
let us represent the spacetime as a manifold. Then, $N$, being a scalar, is
expressible as an integral over the curvature scalars. Indeed, in four
dimensions, \cite{Gilkey}:
\begin{eqnarray}
N & = & \frac{1}{16\pi^2}\int d^4x~\sqrt{\vert
g\vert}\left\{\frac{\Lambda^2}{2} + \frac{\Lambda}{6} R + \frac{1}{180}(
R^{\mu\nu\rho\epsilon}R_{\mu\nu\rho\epsilon}\right. \nonumber \\
& & \left.- R_{\mu\nu}R^{\mu\nu} + 6 \Delta R -\frac{5}{2} R^2) ~~+~~
O(\Lambda^{-1})\right\} \label{Gilkey-eq}
\end{eqnarray}
The first term yields a cosmological constant. It implies that the average
density of sample points needed for the reconstruction of the field and
spacetime obeys $N/V=\Lambda^2/32\pi^2$, except for corrections due to the
curvature terms. Curvature can therefore be viewed as a spatial modulation of
the average sample density, or density of degrees of freedom. The second term
in $S_{size}$ becomes the Einstein action after we set $\alpha=6\pi/\Lambda$.

We could now represent the spacetimes and fields in the formal partition
function $Z[J] = \int e^{-S_{total}}D[\phi]~D[g]$ as concrete manifolds and
concrete functions on those manifolds. This would yield an unwieldy expression
whose evaluation would be plagued, as usual, by the need to mod out physically
equivalent configurations. Alternatively, the sampling-theoretic view suggests
working with the well-defined partition function
\begin{equation}
Z[J] = \sum_{N=1}^\infty \int D[\phi]\int
D[\lambda]~e^{-Tr\left(\frac{6\pi}{\Lambda} +\frac{1}{2}(\Delta+m^2) \vert\phi
)( \phi\vert +\vert J )( \phi\vert\right)}
\end{equation}
which reads in the eigenbasis of the Laplacian:
\begin{eqnarray}
Z[J] & = & \sum_{N=1}^\infty \frac{1}{N!}\int_0^\Lambda
d\lambda_1...\int_0^\Lambda
d\lambda_N \int_{-\infty}^\infty d\phi_1... \int_{-\infty}^\infty d\phi_N~ \nonumber \\
& &  \exp\left(-\frac{6 \pi N}{\Lambda} \nonumber
-\sum_{i=1}^N\left(\frac{1}{2}(\lambda_i+m^2)\phi^2_i-J_i\phi_i\right)\right)
\end{eqnarray}
\bf Discussion. \rm Note that we here tentatively assumed that it is necessary
to sum over all discrete spectra (up to $\Lambda$), irrespective of what type
of Riemannian manifold, if any, they correspond to. Also, in the full theory,
there could be additional terms that induce the cutoff dynamically, for example
\cite{ak-prl-2004}, (counter-) terms that form a power series in the Laplacian,
$( \phi\vert \sum_r c_r \Delta^r\vert\phi )= Tr\left(\sum_r c_r
\Delta^r\vert\phi )( \phi\vert\right)$, whose radius of convergence is
$\Lambda$. This removes fields that contain components beyond the cutoff by
letting their Boltzmann factor vanish. In general, the terms responsible for
the cutoff may of course not be quadratic in the fields. Further, it may be
necessary to handle manifolds with infinite volume and a continuous spectrum.
In this case, one may have to suitably re-scale $\Delta_c$ as
$N\rightarrow\infty$, to keep its spectrum discrete and therefore indicative of
the manifold's shape.

Since any candidate quantum gravity theory must recover QFT for sufficiently
large length scales, we discussed how the sampling theoretic natural UV cutoff
would impact the QFT path integral. The new framework for the sampling and
reconstruction of spacetimes and fields could be useful, however, also beyond
QFT in various studies of quantum gravity, in particular, in studies in which
spacetime is modelled as discrete, see {\it e.g.} \cite{discreteqg}. There, the
new sampling methods could be used to give discrete structures a continuous
representation. This could establish and stabilize the effective dimension of a
lattice and it could make it unnecessary to take a continuum limit. It could
also serve as a mere mathematical tool in these theories, for example, to
rewrite hard-to-sum series as integrals, or to rewrite hard-to-solve finite
difference equations as differential equations that are easier to handle. Work
is in progress on linking manifold and graph Laplacians, and also on
reconstructing the shape of Lorentzian manifolds from the mutual distances of
events as measured through 2-point functions. Also, formulations of general
relativity in terms of the eigenvalues of the Dirac operator have been
discussed, {\it e.g.}, in \cite{conneslandirovelli}. It should be interesting
to transfer results of those works into the information theoretic framework
here.

\bf Acknowledgement. \rm This work has been supported by the Discovery and CRC
Programs of NSERC. Thanks to K.-H. Rehren for helpful comments.

\end{document}